\documentclass[12pt]{amsart}

\usepackage{amsaddr}
\usepackage[authoryear,round]{natbib}
\usepackage{float}
\usepackage{hyperref}
\usepackage[pdftex]{graphicx}
\usepackage{caption}
\usepackage[a4paper, margin=2cm]{geometry}
\usepackage{bm}

\newtheorem{prop}{Proposition}
\theoremstyle{remark}\newtheorem{rmk}{Remark}

\begin{document}

\title[Statistical Properties of Cohen's d]{Some Additional Remarks on Statistical Properties of Cohen's d from Linear Regression}

\author[J.~Gro{\ss}]{J\"{u}rgen Gro{\ss}{$^{1}$}}
\address{{$^{1}$}Institute for Mathematics and Applied Informatics, University of Hildesheim, Germany}
\email{$^1$juergen.gross@uni-hildesheim.de}
\author[A. M\"{o}ller]{Annette M\"{o}ller{$^{2,}$}{$^3$}}
\address{$^{2}$Faculty of Business Administration and Economics,  Bielefeld University, Germany,\\
$^{3}$Helmholtz Centre for Infection Research (HZI), Braunschweig, Germany}
\email{$^2$annette.moeller@uni-bielefeld.de}
\email{$^3$annette.moeller@helmholtz-hzi.de}
\thanks{{$^{2,}$}{$^3$}Support by the Helmholtz Association’s pilot project ”Uncertainty Quantification” is gratefully acknowledged.}



\date{}

\begin{abstract} The size of the effect of the difference in two groups with respect to a variable of interest may be estimated by the classical Cohen's $d$. A recently proposed generalized estimator allows conditioning on further independent variables within the framework of a linear regression model. In this note, it is demonstrated how unbiased estimation of the effect size parameter together with a corresponding standard error may be obtained based on the non-central $t$ distribution. The portrayed estimator may be considered as a natural generalization of the unbiased Hedges' $g$. In addition, confidence interval estimation for the unknown parameter is demonstrated by applying the so-called inversion confidence interval principle. The regarded properties collapse to already known ones in case of absence of any additional independent variables. The stated remarks are illustrated with a publicly available data set.
\end{abstract}

\maketitle

\section{Introduction}\label{sec:intro}

Consider independent samples of sizes $n_{0}$ and $n_{1}$ of a statistical variable $Y$ in two groups such that
$Y$ follows a normal distribution with expectation $\mu_{0}$ and variance $\sigma^2$ in group ''0'' and expectation $\mu_{1}$ and the same variance $\sigma^2$ in group ''1''.  As an estimator for the size $d$
of the group effect, \citet[p. 66ff]{cohen1988statistical} considers
\begin{equation}\label{E1}
\widehat{d} = \frac{\overline{Y}_{1}-\overline{Y}_{0}}{S}, \quad S =\sqrt{\frac{Q_{0} + Q_{1}}{n_{0} + n_{1} - 2}}\; ,
\end{equation}
where $\overline{Y}_{0}$ and $\overline{Y}_{1}$ are the sample means and $Q_{0}$ and $Q_{1}$
are the sums of squared differences from the respective sample means in the two groups. The estimator $\widehat{d}$
is related to the test statistic $t$ of the usual two-sample test statistic for the null hypotheses
$H_{0}: \mu_{0} = \mu_{1}$ versus the alternative $H_{1}: \mu_{0} \not = \mu_{1}$
by the formula
\begin{equation}\label{E2}
\widehat{d} = t \, \sqrt{\frac{n_{0}+ n_{1}}{n_{0} n_{1}}}\; ,
\end{equation}
see (2.5.3) in \citet{cohen1988statistical}.
According to \citeauthor{cohen1988statistical}, values $|\widehat{d}| =0.2$, $|\widehat{d}|=0.5$ and $|\widehat{d}|=0.8$ indicate a small, medium and large effect, respectively.

When the variable $Y$ of interest is considered to possibly depend on a number of explanatory variables, one may consider a linear regression model described by
\begin{equation}\label{E3}
Y = \beta_{0} + \beta_{1} X_{1} + \beta_{2} X_{2} + \cdots + \beta_{1+k} X_{1+k} + \varepsilon,
\end{equation}
where $\varepsilon$ follows a normal distribution with expectation $0$ and variance $\sigma^2 >0$.
The grouping variable $X_{1}$ takes $0$ as a value when a a response observation falls into group "0" and $1$ when an observation falls into group "1". There are $k$ further explanatory variables $X_{2}, \ldots, X_{1+k}$, which are assumed as absent in the simple case $k=0$.

From equation (\ref{E3}) the expectation of $Y$ conditional on the group is given as \begin{equation}
\mu_{0} = \text{E}[Y|X_{1} = 0] = \beta_{0} + \beta_{2} X_{2} + \cdots + \beta_{1+k} X_{1 + k}
\end{equation} in group ''0'' and $\mu_{1} = \text{E}[Y|X_{1} = 0] + \beta_{1}$ in group ''1''.
Hence,
\begin{equation}\label{E6}
d = \frac{\mu_{1} - \mu_{0}}{\sigma}  = \frac{\beta_{1}}{\sigma}
\end{equation}
is the unknown population effect size, see \citet[(2.5.1)]{cohen1988statistical}.

Recently, \citet{gross2023note} considered a natural generalization of Cohen's estimator from (\ref{E2}) based on the above outlined properties in the regression setting with additional explanatory variables. This estimator is given by
\begin{equation}\label{E7}
\widehat{d} = \frac{\widehat{\beta}_{1}}{\sqrt{\widehat{\sigma}^2}}\; ,
\end{equation}
where $\widehat{\beta}_{1}$ is the least squares estimator of $\beta_{1}$ and $\widehat{\sigma}^2$ is the usual unbiased estimator for $\sigma^2$ under model (\ref{E3}).
In matrix notation the model may also be written  as
\begin{equation}\label{E4}
\bm{Y} = \bm{X} \bm{\beta}  + \bm{\varepsilon}
\end{equation}
where $\bm{Y}$ represents the $n \times 1$ vector of observations of the explained variable $Y$, the $n\times (2 +k)$ model matrix $\bm{X}$ is assumed to be of full column rank, and $\bm{\varepsilon}$ follows a $n$-variate normal distribution with
expectation vector $\bm{0}$ and variance-covariance matrix $\sigma^2 \bm{I}_{n}$ with $\bm{I}_{n}$ denoting the $n\times n$ identity matrix. Then
\begin{equation}
\widehat{\beta}_{1} = \bm{e}^{\prime} \widehat{\bm{\beta}}, \quad \widehat{\bm{\beta}} = (\bm{X}^{\prime} \bm{X})^{-1} \bm{X}^{\prime} \bm{Y}\; ,
\end{equation}
where $\bm{e}$ is a $n\times 1$ vector of {$0$}s except for a $1$ at the position of $\beta_{1}$ in the $(2+k)\times 1$ parameter vector $\bm{\beta} =(\beta_{0}, \beta_{1}, \beta_{2}, \ldots, \beta_{k})^{\prime}$ with $\beta_{2}$ up to $\beta_{k}$ considered as absent in case $k=0$. Moreover,
$$
\widehat{\sigma}^{2} = m^{-1} (\bm{Y} - \bm{X} \widehat{\bm{\beta}})^{\prime}(\bm{Y} - \bm{X} \widehat{\bm{\beta}}), \quad m =  n - 2 - k\; .
$$
In this setting, as noted by \citet{gross2023note}, both formulas
(\ref{E7}) and (\ref{E2}) yield identical estimates for the special case of $k=0$.

\section{Statistical Properties of Cohen's d}

Consider the two quantities
\begin{equation}\label{E10}
v_{1}^2 = \sigma^{-2} \, \text{var}(\widehat{\beta}_{1}) = \bm{e}^{\prime} (\bm{X}^{\prime} \bm{X})^{-1} \bm{e}
\quad \text{and}\quad \tau = \frac{\beta_{1}}{\sqrt{\sigma^2 v_{1}^{2}}} = \frac{d}{\sqrt{v_{1}^{2}}}\; .
\end{equation}
For example from \citet[Theorem 3.5]{seber2003linear}, it is seen that the random variable $X = \widehat{\beta}_{1}/\sqrt{\sigma^2 v_{1}^{2}}$ follows a normal distribution with expectation
$\tau$ and variance $1$, and the random variable $Y =  m \widehat{\sigma}^{2}/\sigma^2$ independently follows a (central) chi squared distribution with $m$ degrees of freedom. Then, the ratio $X/\sqrt{Y/m} = \widehat{d}/\sqrt{v_{1}^{2}}$ follows a non-central $t$ distribution, see e.g. \citet{johnson1940applications}. Thus, one may state the following.

\begin{prop}\label{P1} Under the assumptions of model (\ref{E4}),
\begin{equation}
\frac{\widehat{d}}{\sqrt{v_{1}^{2}}} \sim t(m, \tau)\; ,
\end{equation}
where $t(m, \tau)$ denotes the non-central $t$ distribution with $m$ degrees of freedom and non-centrality parameter $\tau$.
\end{prop}

In case $k=0$ the model matrix $\bm{X}$ becomes
\begin{equation}
\bm{X} = \begin{pmatrix}
\bm{1}_{n_{0}} & \bm{0}\\
\bm{1}_{n_{1}} & \bm{1}_{n_{1}}
\end{pmatrix}\; ,
\end{equation}
where $\bm{1}_{\nu}$ denotes the $\nu\times 1$ vectors of {$1$}s, and a little matrix algebra reveals
\begin{equation}\label{E9}
v_{1}^{2} = \frac{n_{0} + n_{1}}{n_{0} n_{1}}\; .
\end{equation}
Then Proposition \ref{P1} is easily seen to reduce to the result given by \citet[Sect. 3]{hedges1981distribution}.

From \citet{johnson1940applications}, the expectation and variance of the $t(m, \tau)$ distribution are given as
\begin{equation}\label{E8}
\mu_{1}  = c(m) \tau \quad \text{and}\quad \mu_{2} = \frac{m}{m-2} + \left(\frac{m}{m-2} - c(m)^2\right) \tau^2\; ,
\end{equation}
where
\begin{equation}
c(m) = \frac{\Gamma((m-1)/2)\sqrt{m/2}}{\Gamma(m/2)}\; .
\end{equation}

\begin{rmk}\label{R1}
From \citet{tricomi1951asymptotic} one may conclude
\begin{equation}
c(m) = 1  + \frac{3}{4 m} + O(m^{-2}) \quad \text{as $m \rightarrow \infty$}\; ,
\end{equation}
implying that $\lim_{m\to \infty} c(m) = 1$.
\end{rmk}

From Remark \ref{R1}, the number $1  + 3/(4 m)$ may serve as a rough approximation of
$c(m)$, which, however, is less precise than the proposal
\begin{equation}\label{E11}
c(m) \approx \left(1- \frac{3}{4m-1}\right)^{-1}
\end{equation}
from \citet{hedges1981distribution}, see also Table 2 in \citet{goulet2018review} for a comparison of exact values with corresponding approximations. As another
approximation not further investigated here one may consider $c(m)\approx \sqrt{2 m/(2m-3)}$ for larger $m$, which may be concluded from Theorem 2.1 in \citet{laforgia2012asymptotic}.

Now, combining Proposition \ref{P1} with (\ref{E8}) and noting $d = \tau \sqrt{v_{1}^{2}}$ gives the following.

\begin{prop}\label{P2} Under the assumptions of model (\ref{E4}) with $m = n - 2 - k > 2$,
\begin{equation}
\text{E}(\widehat{d}) = c(m) d \quad \text{and} \quad
\text{Var}(\widehat{d}) = \frac{m }{m-2} v_{1}^{2} + \left(\frac{m}{m-2}- c(m)^2\right)  d^2 \;
\end{equation}
for $\widehat{d}$ from \eqref{E7}.
\end{prop}

When considering the asymptotic behaviour of $\widehat{d}$ for increasing number of observations it is reasonable to assume that the group size proportion remains constant, i.e.
\begin{equation}
n_{0} =  \gamma n\quad \text{and}\quad n_{1} = (1-\gamma) n
\end{equation}
for some $0 < \gamma <1$ and any positive integer $n$. Then, letting $n$ approach $\infty$ is equivalent to letting $m$ approach $\infty$, provided the number $k$ of additional independent variables does not depend on the number of observations.

\begin{rmk} From the above Proposition \ref{P2} and Remark \ref{R1}, it follows that $\widehat{d}$ is asymptotically unbiased for $d$, i.e. $\lim_{m\to \infty} \text{E}(\widehat{d}) = d$. If $\lim_{m\to \infty} \text{var}(\widehat{\beta}_{1}) = 0$, then $\lim_{m\to \infty} \text{var}(\widehat{d}) = 0$, in which case $\widehat{d}$ is consistent in  quadratic mean for $d$ under model (\ref{E4}).
\end{rmk}

From (\ref{E9}) it is easily seen that the assumption $\lim_{m\to \infty} \text{var}(\widehat{\beta}_{1}) = 0$ is satisfied when $k=0$. The consistency of $\widehat{d}$ in this case has already been noted by \citet[p. 112]{hedges1981distribution}.

From Proposition \ref{P2} an unbiased estimator for $d$ is provided by $\widehat{d}_{u} = c(m)^{-1} \widehat{d}$, which has also been called Hedges' $g$ when $k=0$, cf. \citet{hedges1981distribution}. A corresponding standard error may be derived from Proposition \ref{P2} by considering the square root of $\text{var}(\widehat{d}_{u})$ with $d$ replaced by $\widehat{d}_{u}$.

\begin{rmk}\label{R2} An unbiased estimator for the parameter $d$ is given by $\widehat{d}_{u} = c(m)^{-1} \widehat{d}$ with corresponding standard error
\begin{equation}
\text{se}(\widehat{d}_{u}) = \sqrt{\frac{m\, c(m)^{-2}}{m-2} v_{1}^{2} + \left(\frac{m\, c(m)^{-2}}{m-2}- 1\right)  {\widehat{d}_{u}}^2}\; .
\end{equation}
\end{rmk}

As also noted in \citet{goulet2018review}, unbiased estimation is to be preferred over biased estimation, but for large $m$ the difference between $\widehat{d}$ and $\widehat{d}_{u}$ is quite small.

From Proposition \ref{P1} it is possible to construct a confidence interval for the parameter $\tau$ defined in (\ref{E10}) by applying the inversion confidence interval principle from  Proposition 2 in \citet{steiger1997noncentrality}. For this, let $F(\tau) \equiv \text{Pr}((-\infty, \widehat{d}/\sqrt{v_{1}^{2}}])$ be the cumulative distribution function of the $t(m,\tau)$ distribution with $m=n-2 - k$ degrees of freedom at $\widehat{d}/\sqrt{v_{1}^{2}}$, considered as a function of the non-centrality parameter $\tau$. For a specified $\alpha$ with $0< \alpha < 1$ let $\tau_{1}$ satisfy
$F(\tau_{1}) = 1-\alpha/2$ and let $\tau_{2}$ satisfy $F(\tau_{2}) = \alpha/2$. Then the interval $[\tau_{1}, \tau_{2}]$ specifies a $(1-\alpha)$ confidence interval for $\tau$. Hence we may state the following.

\begin{rmk} Let $\tau_{1}$ and $\tau_{2}$ be obtained as described above. Then the interval
$$
\left[\tau_{1}\sqrt{v_{1}^{2}},\,  \tau_{2} \sqrt{v_{1}^{2}}\right]
$$
specifies a $(1-\alpha)$ confidence interval for the parameter $d$.
\end{rmk}

Note that this approach has also been illustrated in Example 3 by \citet{steiger1997noncentrality} for the special case $k=0$.

\section{Data Example}

To illustrate and apply the above listed properties, a data set available from the UCI machine learning repository is employed, see \citet{Dua:2019}. It contains student achievement measurements in secondary education of two Portuguese schools, see \citet{cortez2008using}. The following computations are carried out with the  statistical software {\sf R} \citep{Rsoftware}.

\begin{figure}[hbt]
\centering
\includegraphics[width=24pc,angle=0]{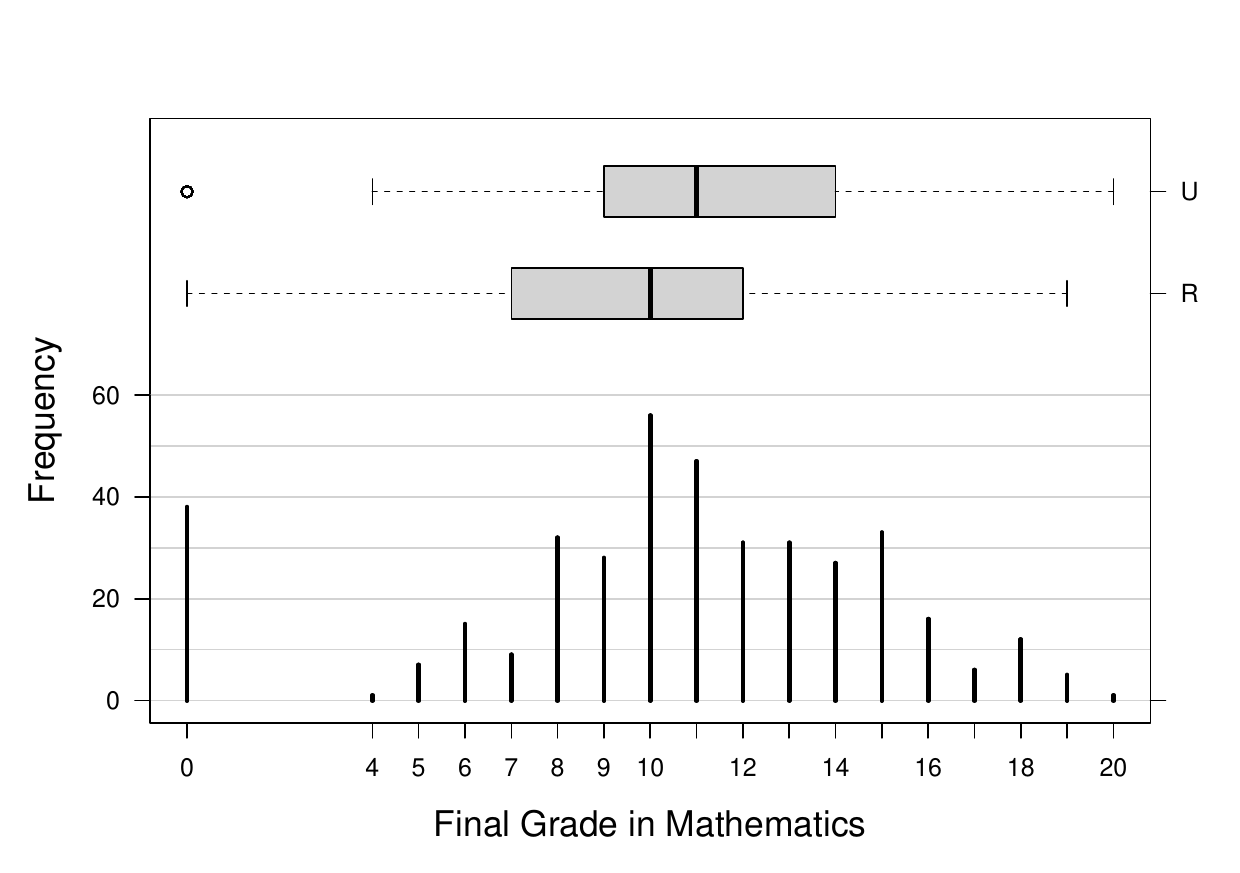}\\
  \caption{Frequency distribution of final Math results (variable G3) from $n=395$ students}\label{F1}
\end{figure}

We consider the final Mathematics grade  with integer values between 0 and 20 as the dependent variable $Y$ from $n=395$ students in two groups. The group $0$ is defined by home address indicated as `rural' with $n_{0} = 307$ observations, while group $1$ is
defined by home address indicated as `urban' with $n_{1} =88$ observations, see Figure \ref{F1}. On average, students from group 1 perform better ($\overline{Y}_{1} = 10.674267$) than students form group $0$ ($\overline{Y}_{0} = 9.511364$).
The corresponding  two-sample two-sided $t$ test statistic with equal variances reads $|t| = 2.1084$  admitting a p-value of $0.03563$. Hence, one may conclude a significant  difference at significance level 0.05.

Cohen's $d$ estimator
from (\ref{E2}) reads
\begin{equation}
\widehat{d} = 2.1084 \sqrt{\frac{307 + 88}{307 \cdot 88}} = 0.2549\; ,
\end{equation}
indicating a rather small effect. The very same value may also be obtained from the {\sf R} package {\tt effectsize}, see \citet{R:effectsize}, except for an opposite sign. This is supposed to originate from using the difference $\overline{Y}_{0} - \overline{Y}_{1}$ instead of $\overline{Y}_{1} - \overline{Y}_{0}$ in the involved formulas. As a matter of fact the $t$ test statistic in {\sf R} is computed by applying the first difference, while the regression coefficient $\widehat{\beta}_{1}$ is identical to the second difference for the special case of $k=0$. In the general regression context with $k>0$ the coefficient $\widehat{\beta}_{1}$ is the estimated positive or negative increment of the intercept in group 1 compared to group 0 conditional on the independent variables -- and may even receive an opposite sign to the unconditional mean difference $\overline{Y}_{1} - \overline{Y}_{0}$.

From fitting
the model $Y= \beta_{0} + \beta_{1} X_{1} + \varepsilon$ one gets $\widehat{\beta}_{1} = 1.162903$ and
$\sqrt{\widehat{\sigma}^{2}} = 4.561543$ yielding the same estimate $\widehat{d}$ by (\ref{E7}) as before.
By using the approximation $c(n-2) \approx 1.001913$ from (\ref{E11})
one gets $\widehat{d}_{u} = c(n-2) \widehat{d} = 0.2544$. Except for the sign, this is also exactly the value of Hedges' $g$ computed from the package {\tt effectsize}.

Now two additional independent variables are considered. The home to school travel time (incorporated as numeric variable $X_{2}$ with values 1  to 4 corresponding to travel times less that 15 minutes, 15 to 30 minutes, 30 to 60 minutes and more than 60 minutes) and the number of past class failures (incorporated as numeric variable $X_{3}$ with values from $0$ to $4$ where $4$ is noted for more than 3 failures). Then from fitting a model
\begin{equation}
Y = \beta_{0} + \beta_{1} X_{1} + \beta_{2} X_{2} + \beta_{3} X_{3} + \varepsilon
\end{equation}
one gets
\begin{equation}
\widehat{\beta}_{1} = 0.6212941,\quad   v_{1}^{2} = 0.01642807,\quad  \sqrt{\widehat{\sigma}^2} = 4.265321\; .
\end{equation}
As noted by \citet{gross2023note}, the estimated regression coefficient $\widehat{\beta}_{1}$ is also identical to the group mean difference $\overline{Y}_{\ast 1}  - \overline{Y}_{\ast 0}$ when a new variable
$Y_{\ast} =  Y - \widehat{\beta}_{2}X_{2} - \widehat{\beta}_{3}X_{3}$ is created. This does, however, not imply that the classical effect size formulas may simply be applied with $Y$ replaced by $Y_{\ast}$, since that would ignore an additional required adjustment for the degrees of freedom, see formula (4) in \citet{gross2023note}.
The (biased) effect size estimate then reads
\begin{equation}
\widehat{d} = \frac{\widehat{\beta}_{1}}{\sqrt{\widehat{\sigma}^2}} = 0.1456617
\end{equation}
implying a very small net group effect size when home school travel time and number of past failures are held constant.
As noted by \citet{gross2023note}, this value may also be converted to Cohen's $f^2$ as
\begin{equation}
\widehat{f}^2 = \frac{{\widehat{d}}^{2}}{m\, v_{1}^{2}} = 0.003303145\; ,\quad m = n -2 -k = 391\; ,
\end{equation}
being in line with the indication of a very small effect size. The approximation (\ref{E11}) gives $c(m)\approx 1.001923$. Then, by Remark \ref{R2}
\begin{equation}
\widehat{d}_{u} = 0.1453822, \quad \text{se}(\widehat{d}_{u}) = 0.1283604\; .
\end{equation}
The usual approximate normal $95\%$ confidence interval $\widehat{d}_{u} \pm 1.96 \,\text{se}(\widehat{d}_{u})$
reads
\begin{equation}
[-0.1062043, \, 0.3969686]\; .
\end{equation}
To obtain a 95\% confidence interval by the inversion  principle, the cumulative distribution function $F(\tau)$ of the $t(m, \tau)$  distribution with $m=391$ is considered at $\widehat{d}/\sqrt{v_{1}^{2}} = 1.136455$.  Then
$F(\tau_{1}) = 0.975$ for $\tau_{1} = -0.8258511$ and $F(\tau_{2}) = 0.025$ for $\tau_{2} = 3.0973105$. Remark 4 yields
\begin{equation}
[-0.1058510, \, 0.3969886]
\end{equation}
as the corresponding $95\%$ confidence interval for $d$,  quite similar to the above.

\section*{Summary}

As illustrated above, a linear regression model may be applied to obtain the net effect size of a group difference with respect to a response variable of interest when other variables are held constant. The size of the group difference effect with respect to each of the incorporated additional variables, however, naturally remains unaccounted for. Nonetheless the above remarks show that some classical results on Cohen's $d$ carry over within a more general regression framework by (a) applying the estimator from \eqref{E7}, (b) replacing the number of the degrees of freedom $n-2$ by $n-2 -k$, with $k$ being the number of additional independent variables, and (c) replacing the quantity
$(n_{0} + n_{1})/(n_{0} n_{1})$ by $v_{1}^2$, the latter being the diagonal element of the matrix $(\bm{X}^{\prime} \bm{X})^{-1}$ corresponding to the group $1$ regression coefficient. The results fit in between the classical effect size measure for the (unconditional) difference in two groups and the effect size measure $f^2$ usually considered within an even more general regression context.

\end{document}